\documentclass[twocolumn,showpacs,preprintnumbers,amsmath,amssymb]{revtex4}

\usepackage[pdftex]{graphicx}
\usepackage{dcolumn}
\usepackage{bm}

\begin{document}

\title{Density Profiles of Liquid/Vapor Interfaces Away from Their Critical Points}

\author{Wei Bu$^1$, Doseok Kim$^2$ and David Vaknin$^{1*}$}
\affiliation{$^1$Ames Laboratory and Department of Physics and
Astronomy, Iowa State University, Ames, Iowa 50011, USA\\
$^2$Department of Physics, Sogang University, Seoul 121-742, Korea}
\begin{abstract}
We examine the applicability of various model profiles for the liquid/vapor interface by X-ray reflectivities on water and ethanol
and their mixtures at room temperature. Analysis of the X-ray reflecivities using various density profiles shows an error-function like profile is the most adequate within experimental error. Our finding, together with recent observations from simulation studies on liquid surfaces, strongly suggest that the capillary-wave dynamics shapes the interfacial density profile in terms of the error function.
\end{abstract}

\date{\today}

\maketitle

\section{Introduction}
The density profiles of the liquid/vapor interfaces have
been the subject of intense investigations theoretically \cite{Waals1894,Cahn1958,Fisk1969,Buff1965,Weeks1977,Gelfand1990,Stillinger2008}
and experimentally \cite{Beysens1987,Huang1969,Wu1973,Aarts2004} for over  
a century by now.  Near the gas/vapor critical temperature, the thickness of
the interface varies from hundreds to thousands of Angstroms ({\AA})
\cite{Beysens1987,Huang1969,Wu1973}, whereas, it is only a 
few {\AA} thick away from
the critical temperature \cite{Braslau1988,Ocko1994}.  It is by now accepted that the effective thickness of 
the liquid/vapor profile is dominated by capillary waves\cite{Sinha1988,Sanyal1991,Fukuto1998,Fukuto2006} with a smaller contribution from intrinsic 
roughness.   The intrinsic roughness arises from the discreteness of molecules forming the liquid\cite{Vaknin2009}. 
However, the form of the functional density profile at the interface
has not been systematically addressed. 

Several possible density profiles, such as, the hyperbolic tangent function,
based on mean-field theory \cite{Waals1894,Cahn1958} and
the error function, associated with capillary wave theory
\cite{Buff1965}, have been proposed.  Liquid/vapor interface undergoing fluctuation due to thermal capillary wave has been investigated quantitatively by molecular dynamics simulations \cite{Sides99,Ismail06}.
In these studies, two different functions (error- and hyperbolic-tangent-function) were used to fit the interfacial density profile obtained from the simulations.
The surface tension could be calculated from the derivatives of the above functions, which then were compared with the 
surface tension deduced from the pressure difference at the interface.
The above two surface tension values were shown to agree better when the density profile was modeled by the error function than the hyperbolic tangent function. However, a more recent molecular-dynamics (TIP4P-QDP-LJ model) simulations result demonstrates that both the error and hyperbolic tangent functions adequately fit the density profile\cite{Bauer2009}.

Light scattering investigations of the profile close to the critical point have been thoroughly addressed by Huang, Wu, and Webb \cite{Huang1969,Wu1973} who had shown 
that the hyperbolic tangent profile can be excluded based on the analysis of 
critical exponent predictions that favored two closely related profiles (i.e. error function
and the Fisk-Widom profile) \cite{Huang1969,Wu1973}. Although, away from the
critical point, the error function has always been assumed as
the interfacial shape in many X-ray and neutron scattering experiments, there has not been a systematic examination to compare the applicability of other profiles.

Since the liquid/vapor interface of simple liquids is typically a few angstroms wide far from
the critical point, X-ray reflectivity technique is
most suitable to determine the interfacial profile at this length scale. In this manuscript, we report synchrotron X-ray
reflectivities on various liquid/gas interfaces far below the
critical point of two liquids and their mixtures, to examine the applicability of possible profiles that best fit the experimental data.

\section{Theoretical background}
\subsection{Possible density profiles}
X-ray reflectivity probes the electron density (ED) profile along the interfacial normal which, to a good approximation, can be related to the 
molecular density profile.  We will therefore assume that the ED profile represents the density profile from this point on. 
We  label the ED values $\rho_1$ and $\rho_2$ for the electron densities of the gas and the liquid away from the interface, respectively. 
The interfacial profile can then be written, in
density language, as
\begin{equation}\label{ED1}
\rho(z)=\frac{1}{2}\left[(\rho_1+\rho_2)-(\rho_1-\rho_2)f(z)\right],
\end{equation}
where $z$ is the direction normal to the interfacial plane, and
$f$ is a universal monotonic function such that
$f(\pm\infty)=\pm1$. In the present study, $\rho_1\approx 0$ and by denoting $\rho_s=\rho_2$, Eq.\ (\ref{ED1}) can be simplified
as follows
\begin{equation}\label{ED2}
\rho(z)=\frac{\rho_s}{2}(1+f(z)).
\end{equation}

For the ideally flat surface (i.e., zero interfacial
thickness), with step-function-like 
profile (i.e., $f(z)={\rm sign}(z)$ with values $1$ and $-1$ for $z>0$
and $z<0$, respectively, and $(1+f(z))/2\equiv \Theta(z)$), the corresponding X-ray
reflectivity is the so-called Fresnel reflectivity ($R_F$).
The nonzero profile-width results in a departure of
the reflectivity from the Fresnel reflectivity, as discussed below.
\begin{figure}[htl]
\includegraphics[width=3in]{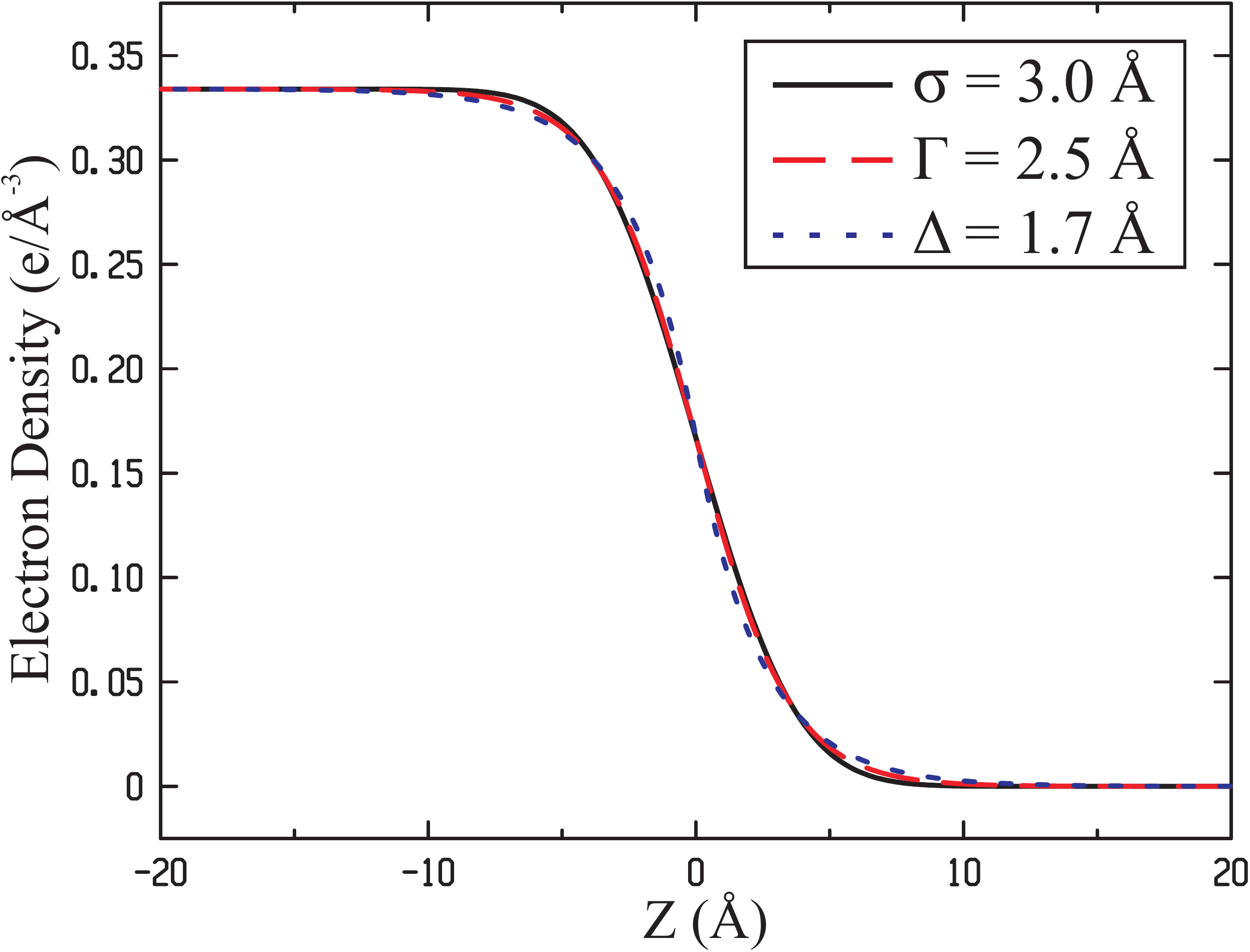}
\caption{\label{sld_mix} (Color online) ED profiles constructed by the error-, hyperbolic-tangent-, and exponential-functions 
of widths $\sigma, \Gamma$, and $\Delta$ respectively, with the same exchange length $L =  3.0$ {\AA}, as discussed in the text. Here, $\rho_s=0.334$ e/{\AA}$^3$ represents the ED of pure water.}
\end{figure}

In the mean-field theory near the gas/liquid critical point the profile is $f(z)=\tanh(\frac{z}{\sqrt{2}\Gamma})$, (TANH), as first derived by van der Waals and Cahn-Hilliard theory \cite{Waals1894,Cahn1958}. Employing capillary wave-theroy,
Buff, Lovett, and Stillinger \cite{Buff1965} predicted the profile of the interface is an error-function
$f(z)={\rm erf}(\frac{z}{\sqrt{2}\sigma})$ (ERF), where $\sigma$ is the
surface roughness.  We compare experimental results to the two profiles, and for further comparison 
also examine the profile produced by the exponential function (EXP) as follows.
\begin{eqnarray}
f(z)&=&\exp(\frac{z}{\sqrt{2}\Delta})/2, \; \; \; z<0 \nonumber
\\
&=&1-\exp(\frac{-z}{\sqrt{2}\Delta})/2, \; z>0
\end{eqnarray}
The roughness parameters
$\sigma$, $\Gamma$, and $\Delta$ of the different profile functions, are all related through an average profile width by the exchange length \cite{Huang1969}, defined by
\begin{equation}\label{exlength}
L=\sqrt{\frac{\pi}{2}}\int^{\infty}_{0}(1-f(z))dz,
\end{equation}
which yields $L=\sigma=\ln2\sqrt{\pi}\Gamma=\sqrt{\pi}\Delta$,
providing a basis for comparing different profile functions.
Figure \ref{sld_mix} shows ED profiles constructed by ERF
(solid line), TANH (dashed line) and EXP (dotted line) with the same exchange length; ($ L = \sigma=3.0$ {\AA}; typical value for the water/vapor
interface at room temperature)\cite{Vaknin2009} , indicating the three line are almost indistinguishable.

\subsection{X-ray Reflectivity}
X-ray reflectivity from an ideally flat surface (step-like function) can be analytically
solved by using Helmholtz equation and the boundary conditions at the interface
$z=0$.  This is usually referred to as the Fresnel reflectivity which is given by
\begin{equation}\label{frensnel}
R_F=\left|\frac{Q_z-Q_s}{Q_z+Q_s}\right|^2,
\end{equation}
where $Q_z$ is the momentum transfer along $z$, and $Q_s=\sqrt{Q^2-16\pi \rho_sr_0}$ ($r_0$ is radius of electron). For $Q_z \gg Q_c$ (above the critical angle for total reflection, $Q_c \equiv \sqrt{16\pi \rho_sr_0}$),
the Fresnel reflectivity (Eq.\ (\ref{frensnel})) can be simplified to $R_F \approx 16\pi^2\rho_s^2r_0^2/Q_z^4$.

For a more structured profile (non-step-like profile), the X-ray reflectivity can be exactly calculated by applying 
 the recursive dynamical method \cite{Parratt1954} by slicing the ED profile \cite{Bu2006} (i.e. ERF, TANH, and EXP). 
Alternatively, the reflectivity can be calculated with high accuracy, especially for  $Q_z \gg Q_c$,
in the Distorted Wave Born approximation (DWBA), such that,
\begin{equation}\label{BA1}
R=R_F\left|\int\frac{d\rho(z)}{dz}\exp(iQ_zz)dz\right|^2,
\end{equation}
Thus, substituting the derivative of the step function, $d\rho(z)/dz=\rho_s\delta(z)$, into Eq. (\ref{BA1}) simply yields the Fresnel Reflectivity $R_F$.  Herein, we apply the exact and the DWBA methods.

\begin{figure}[!]
\includegraphics[width=3in]{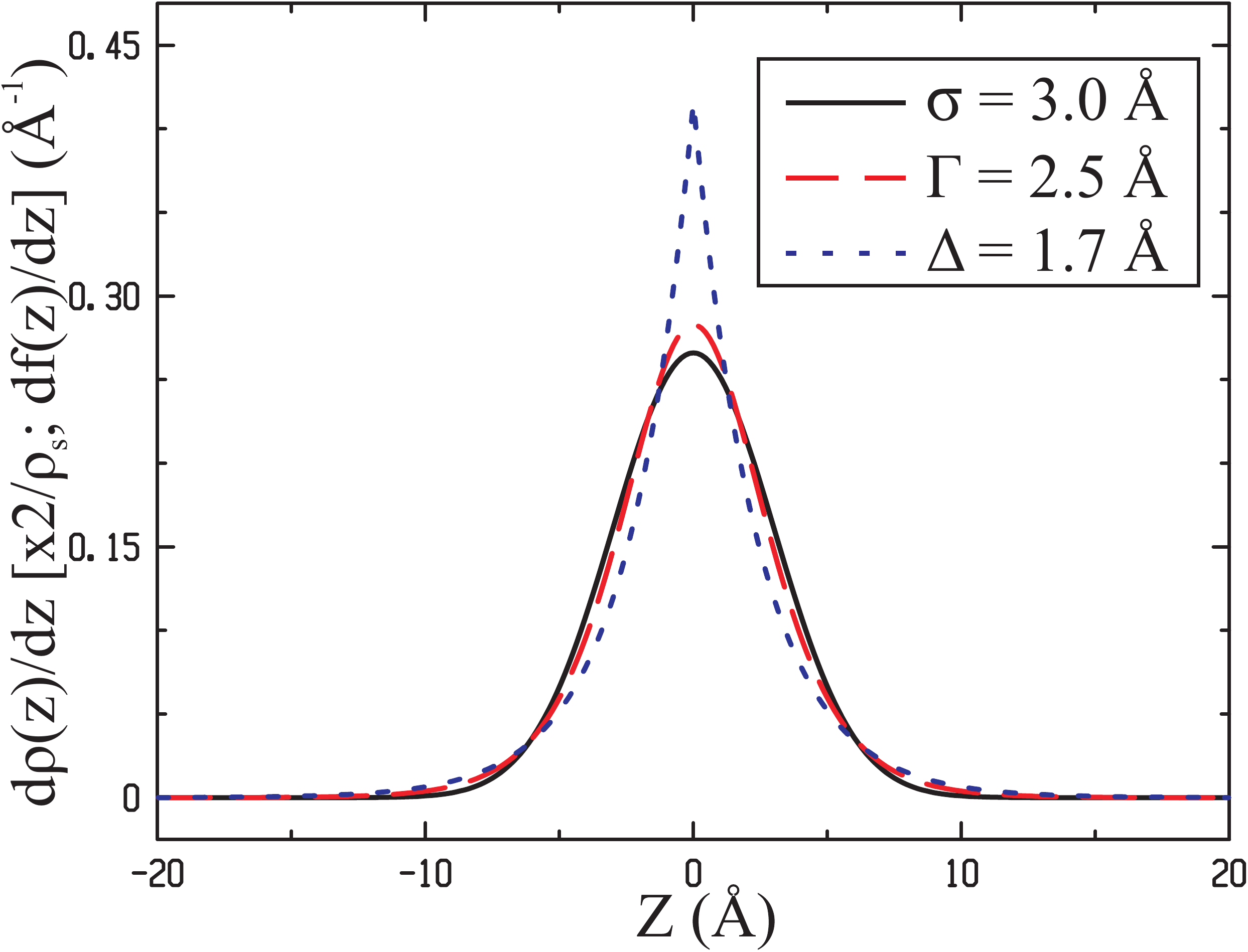}
\caption{\label{dif_mix} (Color online) Derivatives of corresponding ED profiles
shown in Fig. \ref{sld_mix}.}
\end{figure}
As expressed in Eq. (\ref{BA1}) the reflectivity depends on the derivative of the ED profile.
Derivatives of the ED profiles constructed by ERF, TANH, and EXP, are listed as below,
\begin{subequations}
\label{dif}
\begin{eqnarray}
\frac{d\rho(z)}{dz}=\frac{\rho_s}{\sqrt{2\pi}\sigma}\exp(-\frac{z^2}{2\sigma^2}),
\\
\frac{d\rho(z)}{dz}=\frac{\rho_s}{2\sqrt{2}\Gamma}\frac{1}{\cosh^2(\frac{z}{\sqrt{2}\Gamma})},
\\
\frac{d\rho(z)}{dz}=\frac{\rho_s}{2\sqrt{2}\Delta}\exp(\frac{-|z|}{\sqrt{2}\Delta}),
\end{eqnarray}
\end{subequations}
and example calculations are shown in Fig. \ref{dif_mix}.  Although the corresponding ED
profiles shown in Fig. \ref{sld_mix} are hard to distinguish, their derivatives
are quite different, especially for the EXP function.

\begin{figure}[htl]
\includegraphics[width=3in]{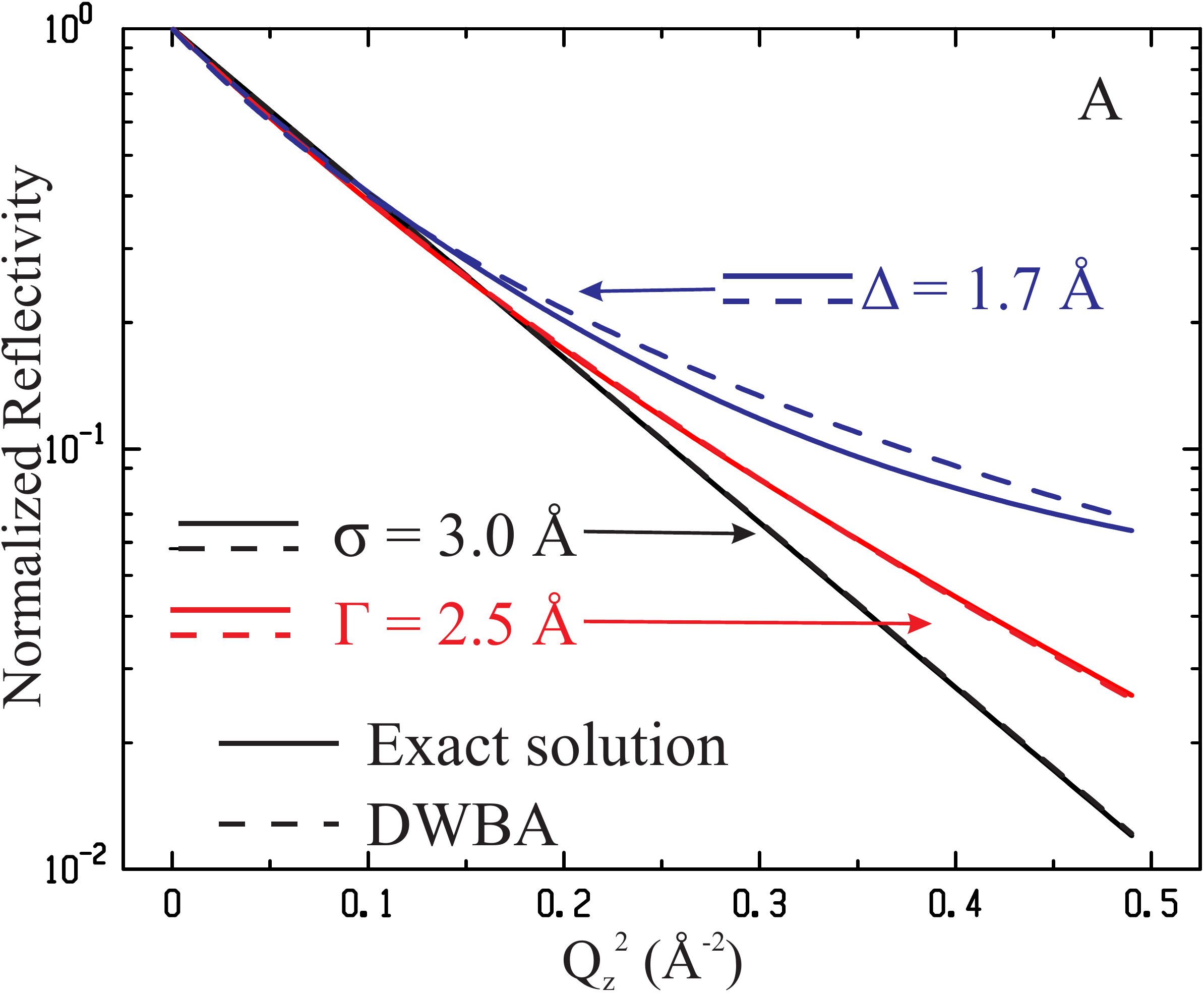}
\includegraphics[width=3in]{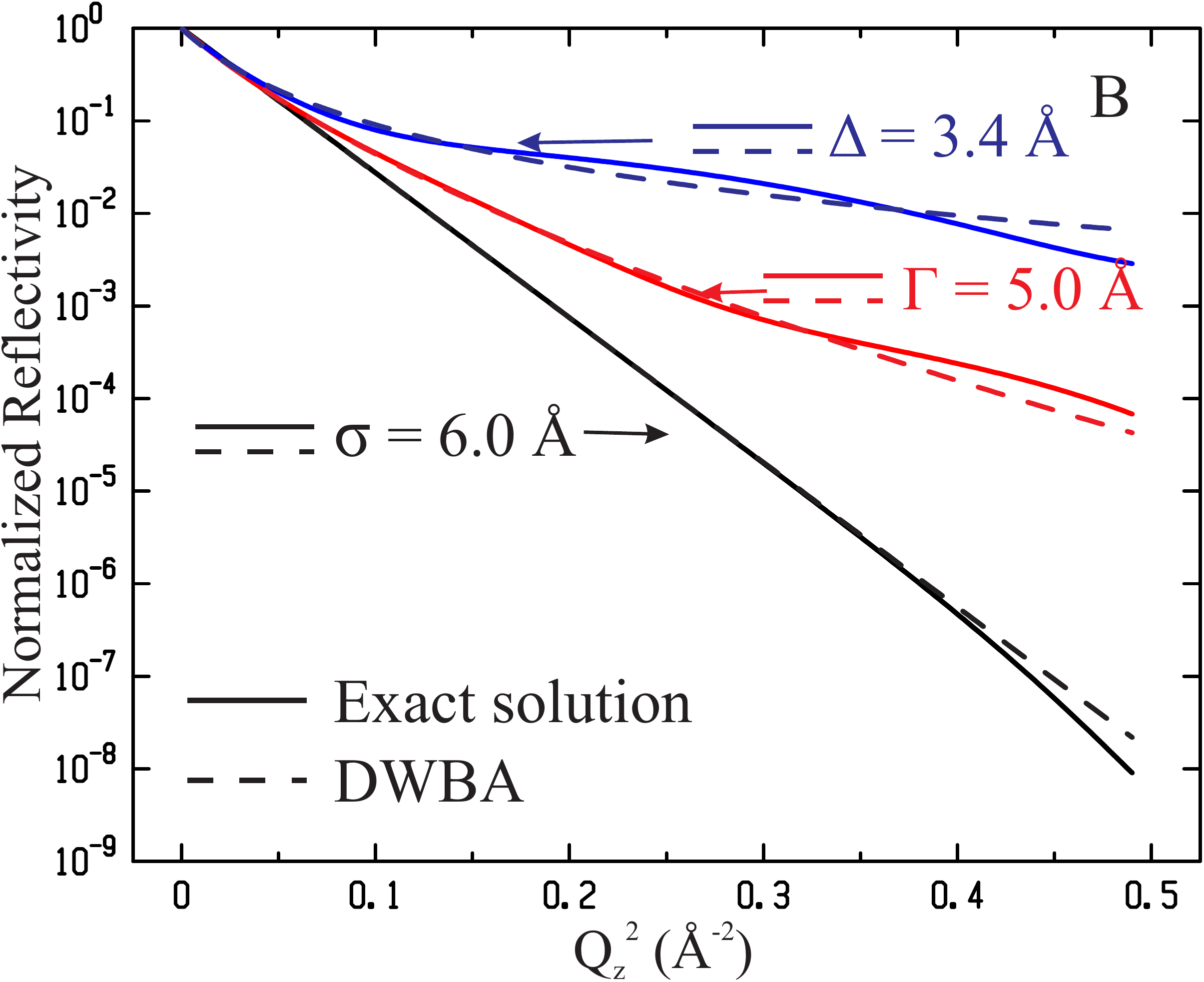}
\caption{\label{rrf_mix} (Color online) Calculated normalized reflectivities of
corresponding ED profiles shown in Fig. \ref{sld_mix} are plotted
{\it versus} $Q_z^2$. Solid lines and dashed lines represent the
exact solution obtained using recursive dynamics method and DWBA, respectively.}
\end{figure}
Substitution of Eq.\ (\ref{dif}) into Eq.\ (\ref{BA1}) and using the
definition of $R_F$ in Eq. (\ref{BA1}) gives the normalized
reflectivities for the three profile functions as listed below:
\begin{subequations}
\label{rrf}
\begin{eqnarray}
\frac{R}{R_F}=\exp(-Q_z^2\sigma^2),\label{rrfa}
\\
\frac{R}{R_F}=\left[\frac{\Gamma
Q_z\pi}{\sqrt{2}\sinh(\frac{\Gamma
Q_\pi}{\sqrt{2}})}\right]^2,\label{rrfb}
\\
\frac{R}{R_F}=\left[\frac{1}{1+2Q_z^2\Delta^2}\right]^2.\label{rrfc}
\end{eqnarray}
\end{subequations}
One should notice that the normalized reflectivities,
$R/R_F$, are independent of subphase ED ($\rho_s$).  Figure
\ref{rrf_mix} shows normalized reflectivities calculated by
recursive dynamical method (solid lines) and Born approximation
(dashed lines) for the corresponding ED profiles shown in Fig.
\ref{sld_mix}. The calculated reflectivities using the two different methods are practically indistinguishable
for ERF and TANH ED profiles when $L=3.0$ {\AA}, whereas the EXP profile differs slightly between the two methods.
As for the density profile, deviation of the EXP profile from the other two is appreciable for large $Q_z$ values.
By contrast,  at larger $L$ (Fig.\ \ref{rrf_mix} for $L = 6 ${\AA}), the three profiles give very different reflectivities even at small 
$Q_z$ values.  Note, that calculated  reflectivities for TANH and
EXP ED profiles are about four to six orders of magnitude larger than the one with
ERF ED profile at large $Q_z$, respectively.

\section{Experimental Details}
The X-ray reflectivity experiments were conducted on the Ames
Laboratory Liquid Surface Diffractometer at the 6ID-B beamline at
the Advanced Photon Source (APS) at Argonne National Laboratory
\cite{Vaknin2012}. The highly monochromatic beam (16.2
keV; $\lambda$ = 0.765335 {\AA}), selected by an initial Si double
crystal monochromator, is deflected onto the liquid surface at a
desired angle of incidence with respect to the liquid surface by a
second monochromator (Ge(220) single crystal), which is placed on
the diffractometer \cite{Vaknin2012}. High beam flux can provide
reflectivity data with good statistics up to $Q_z\sim0.8$
{\AA}$^{-1}$.

Ultrapure water (Millipore, Milli-Q, and NANOpure, Barnstead;
resistivity, 18.1 M$\Omega$cm) and ethanol (99.5 \%, Fisher) were used for all
subphase preparations. Solutions were contained in a stainless
steel trough with the depth of 0.3 mm to reduce the effect of
mechanical agitations on the surface smoothness. To test the
quality of the surface we routinely checked that the reflection
from the surface below the critical angle is nearly 100{\%} (i.e., total reflection). The
trough was encapsulated in an air-tight thermostatic aluminum
enclosure (T = 293 K), which was continuously purged with a helium
gas flow (bubbled through the same mixture as the sample) during
the course of the experiment to lower background scattering from
the air.

\section{Results and Discussions}
\begin{figure}[!]
\includegraphics[width=3in]{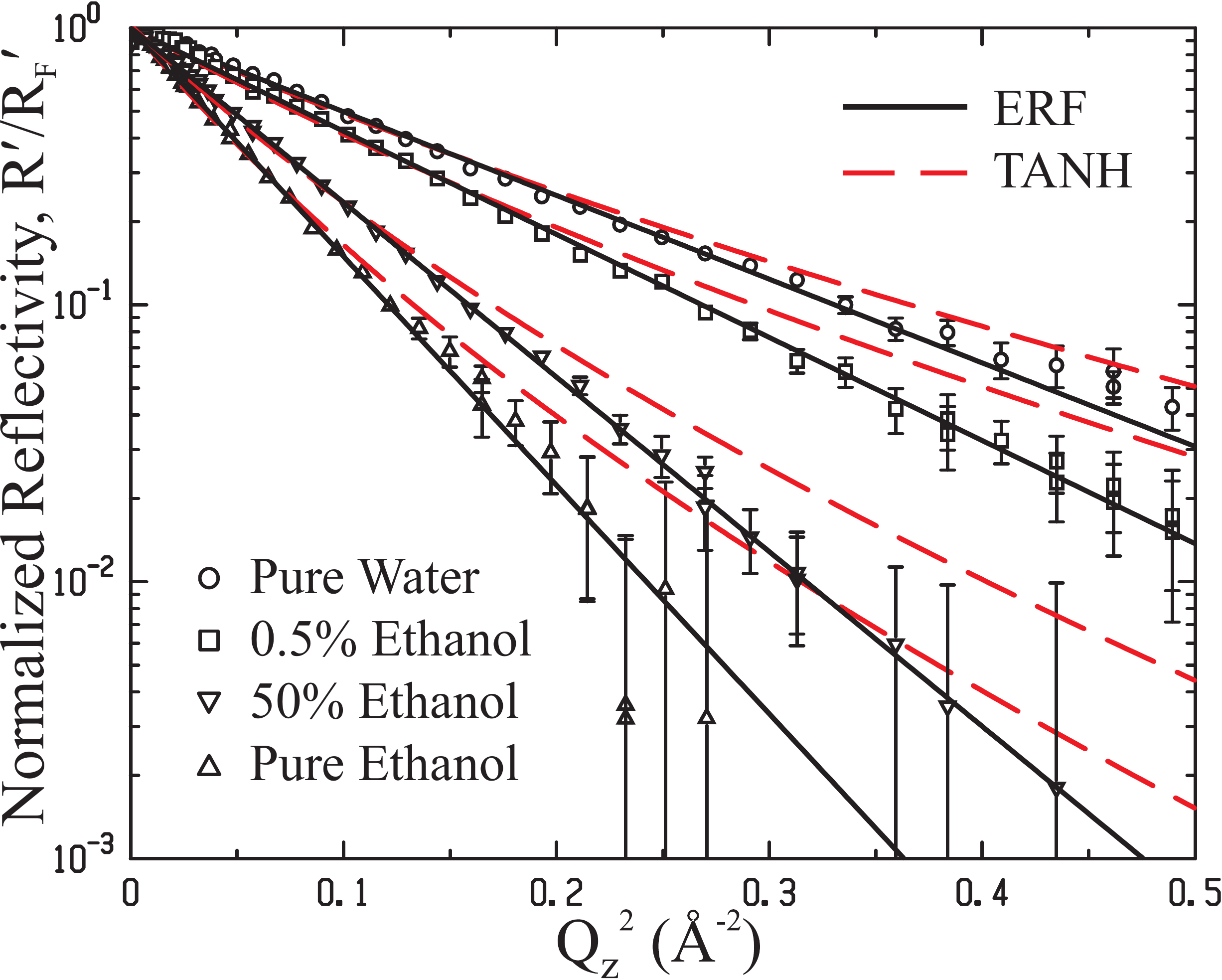}
\caption{\label{exp} (Color online) Normalized reflectivities ($R/R_F$) of pure
water (circles), 0.5{\%} ethanol (squares), 50{\%} ethanol
(inverted triangles), and pure ethanol (triangles). Solid and
dashed lines are the best fits by considering ERF or TANH as ED
profiles, respectively.}
\end{figure}
\begin{table}[htl]
\caption{\label{para} Parameters that generate the best-fit
calculated reflectivities to the experimental data in Fig.
\ref{exp}.}
\begin{ruledtabular}
\begin{tabular}{c|ccc}
subphase& $\sigma$ ({\AA}) & $\Gamma$ ({\AA})& $\sigma/\Gamma$\\
\hline
pure water        &2.64      &2.18    &1.21\\
0.5{\%} ethanol   &2.93      &2.43    &1.20\\
50{\%} ethanol    &3.81      &3.20    &1.19\\
pure ethanol      &4.36      &3.62    &1.21\\
\end{tabular}
\end{ruledtabular}
\end{table}
As discussed above and shown in Fig.\ \ref{rrf_mix}, large roughness ($L>3.0$ {\AA}) is necessary
to test the adequacy of the ERF and TAHN ED
profiles in modeling the measured reflectivities. Pure ethanol at room temperature has much  lower surface tension
(ST; $\sim 23$ mN/m) than that of pure water ($\approx 73$ mN/m), inducing
thermally excited capillary waves with adequate roughness to test the various ED profiles. Thus,
water and ethanol mixtures have been used here for increasing the
roughness gradually by increasing the ethanol portion in the mixture. 
The other apparent way to gradually change interfacial tension is to form and compress a Langmuir monolayer, however such a system is more complicated as it introduces distinct layering at the interface and requires extra fitting parameters\cite{Bu2006}, while the mixture-bare-interface is a simpler surface that can be adequately modeled  by Eq.\ (\ref{rrf}). 
Figure \ref{exp} shows normalized
reflectivities of four different solutions as indicated. Solid and dashed lines
are the best fits calculated by  Eq.\ (\ref{rrfa})
and Eq.\ (\ref{rrfb}), respectively. Calculations by the recursive
dynamical method (exact solutions) are consistent with the DWBA.   
In this process, the only fitting parameter is $\sigma$ or $\Gamma$ 
which are listed in Table \ref{para}. The ratios of $\sigma$ and
$\Gamma$ shown in the forth column of Table \ref{para} are very
close to the criterion ($\sigma/\Gamma=\ln2\sqrt{\pi}\sim1.23$
{\AA}) for both ERF and TANH having the same exchange length, $L$. In
other words, it means that the ED profiles from the fitting results
based on ERF and TANH tend to be as close as possible, in spite
of the fact that the ERF fits the data much better.

In Table \ref{para}, we also note that the interfacial roughness
increases with the mole-fraction of ethanol as  the ST
decreases, consistent with the capillary wave theory. For the
interface with small roughness (i.e., pure water/air), solid (fit based
on ERF) and dashed (fit based on TANH) lines start to diverge away
from each other at $Q_z^2\sim0.2$ {\AA}$^{-2}$ with the rest of the data (Fig.\ \ref{exp}), which
makes almost indistinguishable.   However, as the roughness increases (i.e., pure ethanol), the fit to the ERF (solid lines) is much superior 
than that of the TANH profile (dashed line).   We therefore conclude that the TANH and the EXP profiles 
can be clearly ruled out. 

Our findings compare very well with the simulation studies that show differences between the error function and the hyperbolic tangent function  \cite{Sides99,Ismail06}. The error function was just marginally better than the hyperbolic tangent function\cite{Ismail06} in fitting the interfacial density profile from molecular dynamics simulations, but yielded a sizable difference in the surface tension value such that the error function gave a more satisfactory agreement to the surface tension directly extracted from the pressure difference in the simulations. By contrast, the surface tension based on the hyperbolic tangent function always gave lower estimates. These independent studies on liquid surfaces using X-ray and MD simulations thus strongly suggest that the underlying capillary-wave dynamics dominates the profile of the liquid/vapor interface  for simple liquids, and that the error-function is closest to the average effective interfacial density profile.

\section{Conclusions}
Using  synchrotron X-ray reflectivity techniques on
water/ethanol mixtures surfaces, we find that the interfacial roughness
increases as surface tension decreases as expected from 
 capillary wave theory. Modeling the experimental
data to various ED profiles shows that the error-function profile
is the most adequate within the
experimental uncertainties  for  various surface tension values. 
Although the hyperpbolic-tangent  profile
seems to fit the data well for small momentum transfer  $Q_z$ values, synchrotron measurements that enable large $Q_z$ values clearly indicate the inadequacy  of this profile to the bare liquid/vapor interface.  
We have also considered a hypothetical exponentially decaying profile that totally fails to account for X-ray reflectivity data.  

\acknowledgements {The work at the Ames Laboratory was supported by the Office of Basic Energy Sciences, U.S. Department of Energy under Contract No.  DE-AC02-07CH11358.
 Use of the Advanced Photon Source was supported by the U. S. Department of Energy, Office of Science, Office of Basic Energy Sciences, under Contract No. DE-AC02-06CH11357. D. K was supported by the National Research Foundation (NRF) grant funded by the Korea government (MEST) No. 2011-0017435.}
 
 $^*$email: vaknin@ameslab.gov

\end{document}